# Optical microscopy for pathological imaging: past, present, and future

Runchen Zhang[1,†], Xiaoyi Shang[1,†], Liangyu Deng[1,2], Jingyu Wang[1], Xiaoyu Zheng[3], Lanyu Cui[4], Yao Du[3], Jiaxin Zhuang[3], Yifei Ma[1], Wanying Liu[2], Limei Guo[5], Lin Luo[6], Zhiyi Liu[7], Stephen M. Morris[1], Honghui He[2,*], and Chao He[1,*]

[1]*Department of Engineering Science, University of Oxford, Parks Road, Oxford, OX1 3PJ, UK*
[2]*Institute of Biopharmaceutical and Health Engineering, Tsinghua Shenzhen International Graduate School, Tsinghua University, Shenzhen 518055, China*
[3]*Department of Computer Science and Engineering, The Hong Kong University of Science and Technology, Clear Water Bay, Kowloon, Hong Kong, China*
[4]*School of Basic Medical Sciences, Guangxi Medical University, Nanning, Guangxi 530021, China*
[5]*Department of Pathology, Peking University Third Hospital, School of Basic Medical Sciences, Peking University Health Science Center, Beijing 100191, China*
[6]*College of Engineering, Peking University, Beijing 100871, China*
[7]*State Key Laboratory of Extreme Photonics and Instrumentation, College of Optical Science and Engineering, International Research Center for Advanced Photonics, Zhejiang University, Hangzhou, Zhejiang 310027, China*
[†]These authors contributed equally to this work.
*Corresponding authors: he.honghui@sz.tsinghua.edu.cn; chao.he@eng.ox.ac.uk

## Abstract

**Histopathology, the microscopic examination of stained tissue sections, remains the gold standard for pathology imaging, but its bright-field imaging is largely morphological. Advanced optical microscopy techniques move beyond this by exploiting diverse light–matter interactions, and are increasingly applied to pathological imaging. By capturing signals such as refractive index, polarisation, harmonic generation, Brillouin shift, and molecular vibrations, these microscopy techniques can provide quantitative or semi-quantitative information on tissue structure, mechanics, and chemical or molecular composition. In this review, we summarise the physical principles, key quantitative parameters, and representative pathological applications of seven modalities—optical diffraction tomography, polarisation microscopy, second-harmonic generation microscopy, third-harmonic generation microscopy, Brillouin microscopy, Raman microscopy, and fluorescence microscopy. We also discuss progress in artificial intelligence that supports reconstruction, analysis, and interpretation across these modalities, together with emerging trends in multimodal integration. Overall, this review shows how different optical contrasts can complement routine histopathology, while outlining the challenges and opportunities that will shape their clinical translation.**

## 1. Introduction

Disease processes are commonly assessed through two complementary imaging domains: radiological and pathological imaging. In practice, diagnostic pathways for many major diseases often begin with radiological evaluation using modalities such as ultrasound, X-ray, magnetic resonance imaging, and positron emission tomography (Fig. 1a). These modalities enable early detection and macroscopic localisation of suspicious lesions[1,2], but often lack the specificity needed to resolve histopathological features. Pathological imaging therefore becomes essential once treatment decisions depend on precise tissue classification. The clinical gold standard for pathological imaging remains conventional histopathology, which examines stained tissue sections under a bright-field microscope as an essentially intensity-based, morphological image[3]. Pathologists rely on it for well-established diagnostic criteria, standardised reporting, and a direct fit into everyday surgical workflows[4,5].

This routine bright-field approach enables disease assessment at the cellular and tissue levels, providing the basis for definitive diagnosis, classification, and grading[6,7]. However, it only captures part of the disease-relevant signal such as subtle structural remodelling of the stroma and extracellular matrix (ECM), altered tissue mechanics, and metabolic shifts are difficult to quantify from conventional H&E or special-stained images[8-10]. Advanced optical microscopy addresses these limitations by using properties of light beyond intensity: it converts phase, polarisation, frequency, and their changes in space and time into contrasts in tissue structure, mechanics, and chemical composition (Fig. 1b).

Despite these additional contrasts, most advanced optical microscopy modalities have not yet achieved routine clinical adoption. A persistent barrier is that technical performance is often reported using instrument-centric metrics rather than measures of diagnostic accuracy, prognostic outcomes, or clinical workflow compatibility[11,12]. Thus, for optical microscopy to transition from a research tool to a routine diagnostic adjunct, future development should focus on three things: (1) improving clinical usability, (2) establishing robust links between optical measurements, biological properties and pathological descriptors, and (3) integrating quantitative optical information into existing diagnostic workflows. We note that artificial intelligence (AI) will be important here, converting complex optical measurements into robust, interpretable, and clinically actionable information[4,13-15].

In this review, we focus on seven representative modalities and group them by their dominant pathology-relevant contrast: tissue structure (optical diffraction tomography, polarisation microscopy, and second- and third-harmonic generation microscopy), mechanical properties (Brillouin microscopy), and molecular composition (Raman and fluorescence microscopy) (Fig. 1c). We chose these because they capture the main optical principles behind each contrast, and many other methods are variants or extensions of them. Methods that primarily modify the acquisition, sectioning, or resolution of an existing contrast are discussed within the relevant modality rather than as separate contrast categories. For example, light-sheet[16-18] and super-resolution microscopy[9,19-21] extend the imaging volume and spatial resolution of fluorescence microscopy, while multiphoton excitation[22] increases the imaging depth of both fluorescence and harmonic generation microscopy. Note that modalities whose primary readout is mainly intensity- or reflectance-based tissue morphology, such as confocal[23], optical coherence tomography[24] and optical coherence microscopy[25], fall outside this review.

We organise this review by progressing from optical principles to pathological applications, cross-modality comparison, and outlook. First, we introduce how distinct optical contrasts relate to specific pathological information and provide an overview of representative modalities that access these contrasts. Building on this, we then examine each modality, summarising the imaging principles, key quantitative parameters, representative pathological applications, and the role of AI in supporting clinical translation. Second, we compare representative optical microscopy modalities across five common pathological scenarios, examining how different optical microscopy techniques balance imaging performance, quantitative specificity, interpretability, and workflow compatibility across diverse diagnostic scenarios. Finally, we outline future directions for applying advanced optical microscopy to pathological imaging. Overall, we show how advanced optical microscopy can complement conventional histopathology by adding quantitative or semi-quantitative contrasts that it cannot capture. Crucially, because no single modality is optimal for every task, each pathological task should be matched with the most suitable microscopy modality.

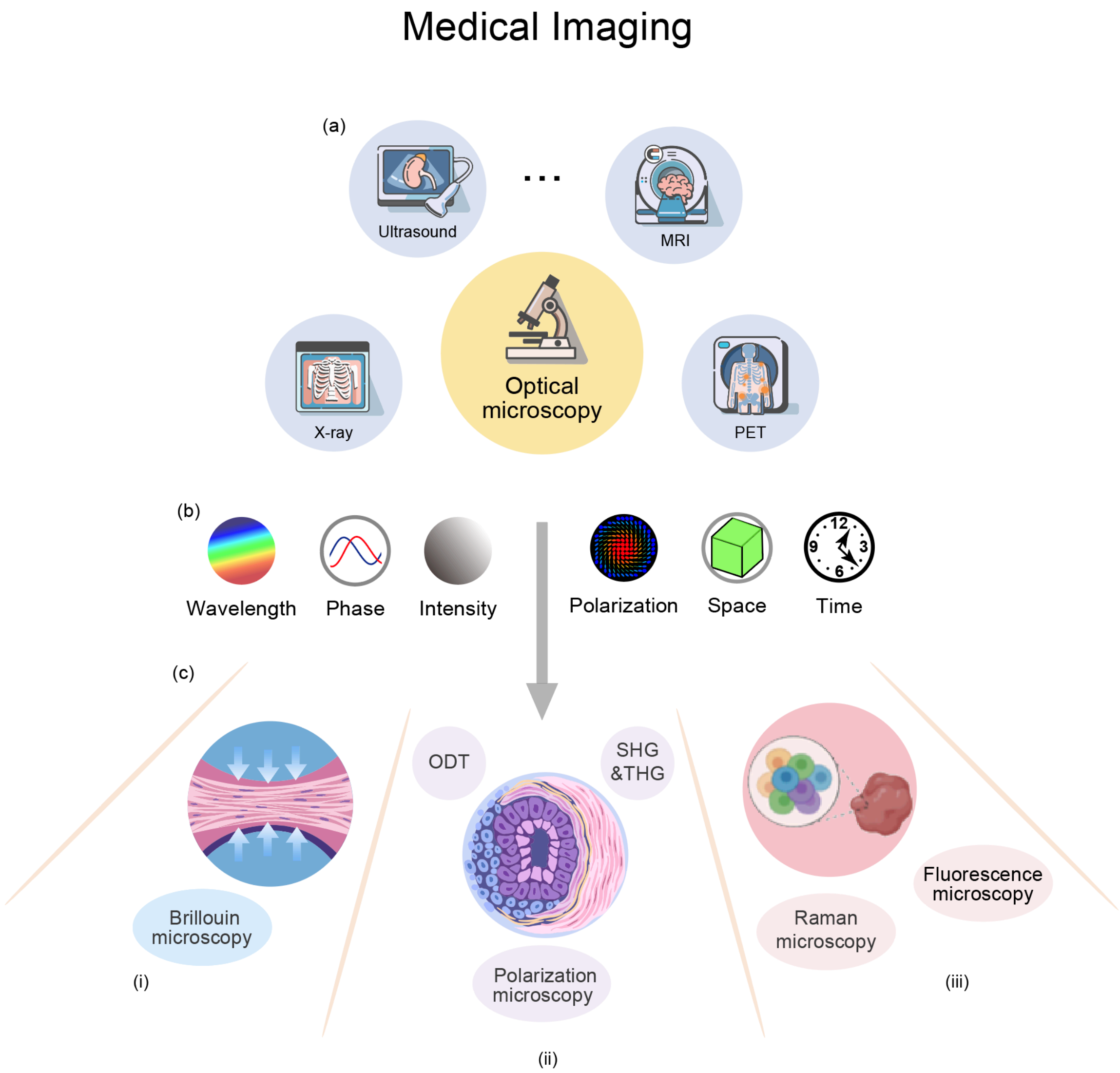

**Fig. 1. Optical microscopy for pathological imaging: scope and organisation.** a, Common medical imaging modalities across two domains: radiological imaging (ultrasound, X-ray, MRI, and PET) and pathological imaging based on optical microscopy. b, Physical properties of light exploited in optical microscopy including intensity, phase, polarisation, frequency, and their changes in space and time[26]. c, Classification of seven representative optical microscopy modalities according to the primary pathological information they provide: (i) mechanical properties, accessed by Brillouin microscopy; (ii) morphological and structural organisation, accessed by ODT, polarisation, SHG, and THG microscopy; and (iii) chemical and molecular specificity, accessed by Raman and fluorescence microscopy. (b) Adapted with permission from ref. [26]. CC BY 4.0.

## 2. Optical contrasts and representative modalities in pathological imaging

### 2.1 Morphological and structural organisation

Pathological diagnosis and classification are grounded in alterations of cellular and tissue-scale morphology, reflecting changes in spatial organisation that integrate genetic, biochemical, and mechanical perturbations[3,27,28]. Optical microscopy provides multiple, complementary ways of accessing this structural information, ranging from volumetric refractive-index (RI) distributions and three-dimensional (3D) cellular geometry to anisotropic fibrous organisation and symmetry-defined ECM architectures. In this section, we focus on three representative contrast families: ODT, polarisation microscopy, and nonlinear harmonic generation, to illustrate how they encode complementary aspects of this structural organisation.

### *2.1.1 ODT*

ODT is a computational tomographic microscopy technique that reconstructs the 3D RI distribution of transparent or weakly scattering specimens, providing quantitative and label-free information on cellular and tissue architecture[29-32]. It does so by combining multiple complex optical field measurements acquired under varying illumination or observation conditions (Fig. 2e). Measurement diversity can be introduced through illumination-angle scanning, specimen rotation, or axial translation, and the resulting measurements are computationally inverted to yield the RI maps[33-41]. Reconstruction strategies range from linear approximations suitable for weakly scattering samples to nonlinear models that partially account for multiple scattering in optically heterogeneous specimens[29,42-47]. Balancing reconstruction fidelity and computational efficiency remains a central challenge, particularly for pathological samples with increased thickness and structural complexity[43,46-48].

Recent technical developments have broadened the scope and improved the robustness of ODT beyond its original formulation[32,46,47]. Polarisation-sensitive and hyperspectral variants add sensitivity to optical anisotropy and wavelength-dependent RI dispersion, expanding the accessible biophysical parameter space[49-51]. Fourier-ptychography-based diffraction tomography improves spatial resolution while relaxing numerical-aperture requirements, and non-interferometric implementations ease coherence constraints and improve compatibility with thick or dynamic samples[52-55]. Such variants make ODT a versatile quantitative imaging modality for biological and pathological applications[32,48].

The pathological value of ODT lies not only in RI contrast itself, but also in the biophysical parameters extracted from reconstructed RI volumes[32,56-59] (Fig. 3a). Morphological descriptors such as cell volume, surface area, and 3D shape can be quantified from tomograms with subcellular spatial resolution[38,56]. Because RI is approximately linearly related to local dry-mass concentration through the specific refractive increment, ODT can estimate cytoplasmic dry-mass concentration, dry-mass density, and total cellular dry mass[57,58]. We note that these estimates assume a relatively uniform specific refractive increment, so they should be interpreted with caution in compositionally heterogeneous tissues[57,60]. Time-resolved ODT measurements additionally capture dynamic membrane fluctuations at millisecond temporal resolution, providing quantitative indicators of cellular mechanical state[59].

In pathological contexts, ODT is particularly useful for studying diseases characterised by subtle yet systematic cellular remodelling that is difficult to quantify using conventional intensity-based histology[32,48]. A representative application is haematological and infectious pathology, where ODT has been applied to red blood cell disorders such as malaria and Babesia infection[56,61-63]. ODT has also been applied in oncology to quantify malignancy-associated phenotypic alterations at the single-cell scale[48,64-67]. Such RI-derived features may reflect cancer-associated changes in cell size and shape, nuclear-to-cytoplasmic ratio, intracellular dry-mass concentration, and hydration state, which together shape local RI distributions. Studies on cells derived from colon cancer, breast cancer, and haematological malignancies have shown that RI-derived parameters, including intracellular mass density and its spatial heterogeneity, are associated with tumour-related phenotypes and intrapopulation variability[64-67].

Despite these advances, extracting reliable and pathologically meaningful information from ODT in optically complex tissues remains a central challenge. Multiple scattering, refractive-index heterogeneity, and model-dependent assumptions degrade reconstruction fidelity, while RI-derived parameters often lack direct biological interpretability[46,47,55]. Learning-based approaches have therefore been introduced at both reconstruction and analysis stages. Data-driven tomographic models mitigate limited-angle and multiple-scattering artefacts by learning inverse mappings from measurements to volumetric RI distributions. At the analysis level, machine-learning algorithms automate extraction and classification of multidimensional phenotypic features from RI volumes, improving robustness and facilitating pathological interpretation[68-70].

### *2.1.2 Polarisation microscopy*

Polarisation microscopy, most comprehensively implemented as Mueller matrix polarimetry, quantifies how biological specimens modify the polarisation state of light[26,71-74] (Fig. 2a). To obtain basic physically interpretable parameters, raw Mueller matrices are typically processed using decomposition or transformation methods that separate retardance, diattenuation, and depolarisation contributions[75]. Note that in complex tissues, these quantities should be regarded as effective polarimetric descriptors whose interpretation depends on the decomposition model and measurement conditions. Advances in optical modulation schemes and polarisation-resolved detection have moved polarisation microscopy from simple crossed-polariser imaging towards faster, quantitative, and increasingly clinically translatable configurations[76-90].

At the physical level, these Mueller-matrix-derived parameters describe distinct optical effects related to tissue microstructure[71,72,86]. Linear retardance reflects the presence and magnitude of birefringent order, for example from aligned collagen fibres or muscle filaments, whereas circular retardance reflects chirality such as that of chiral molecules or helical structures[91-96]. Diattenuation may arise in tissues containing anisotropic scattering structures, and less commonly oriented absorbing components, such as aligned fibrous proteins or elongated cellular architectures; in bulk tissue it is typically weak[97]. Depolarisation reflects the degree of structural heterogeneity and multiple scattering within tissue and is often associated with complex microstructural organisation, irregular interfaces, or increased cellular density[71,98-101]. From these primary parameters, higher-level polarimetric feature parameters, including texture statistics, spatial-heterogeneity indices, and multiscale organisation measures, can be derived as more direct descriptors of tissue remodelling[86,87,94,102-106].

Polarisation microscopy provides semi-quantitative readouts of fibrous tissue organisation, including collagen reorganisation, fibre alignment, and stromal architecture (Fig. 3b). These metrics are particularly valuable for disease staging and progression assessment in fibrotic and desmoplastic conditions[93,94,98,107,108]. First, increased linear retardance and greater spatial uniformity of the optic-axis orientation have been reported in desmoplastic stroma associated with breast, colorectal, liver, and cervical cancers, supporting objective, semi-quantitative assessment of fibrosis severity and stromal activation[93,94,98,107,109]. Second, beyond global stromal characterisation, polarisation microscopy can reveal spatial heterogeneity in tissue organisation. Local variations in retardance, continuity of the optic-axis orientation, and depolarisation can distinguish aligned collagen networks from disordered or strongly scattering regions associated with inflammation or necrosis, enabling structure-based phenotyping of the tissue microenvironment[95,99,100,103]. Third, polarisation contrast may contribute to tumour-margin assessment by detecting early architectural disruption of collagen networks that could precede overt cytological invasion, as suggested in colorectal, cervical, and gastrointestinal studies[107,110,111]. Finally, polarisation contrast is not confined to thin histological sections. In reflectance and endoscopic configurations, birefringence and depolarisation reveal subsurface architectural abnormalities in bulk tissue, supporting real-time intraoperative guidance and in vivo screening[99-101,112-118].

Despite these capabilities, polarisation microscopy faces challenges in data acquisition and diagnostic reliability[26,71,73,119]. On the acquisition side, polarimetric parameters depend on imaging geometry, wavelength, spatial scale, and signal-to-noise conditions, complicating cross-study comparison and reproducibility[71,72,120-128]. Learning-based denoising and calibration methods have been introduced to stabilise measured Mueller matrix elements under rapid or low-signal acquisition conditions[129,130]. Neural-network-based regressors can also accelerate extraction of polarisation parameters when conventional decomposition becomes computationally intensive or noise-sensitive[129,130]. On the diagnostic side, multiple scattering and depolarisation can obscure physically meaningful contrast in heterogeneous tissues and complicate interpretation[71,99-101,131]. Machine-learning approaches can support the characterisation of collagen fibre organisation and classification of pathological tissues using polarisation-derived features[104,132]. Supervised machine-learning and deep-learning models applied to polarimetric parameter sets have demonstrated improved discrimination of pathological tissue states by integrating multiple physical features and capturing complex spatial patterns[129,130,133,134].

### *2.1.3 SHG and THG microscopy*

SHG and THG microscopy are nonlinear optical imaging techniques that provide label-free and structure-specific contrast by exploiting higher-order light–matter interactions[135-137]. SHG is a second-order nonlinear optical process in which two incident photons combine to coherently generate a photon at twice the excitation frequency[138] (Fig 2c). This process is governed by the second-order nonlinear susceptibility and vanishes in centrosymmetric media under the electric-dipole approximation, rendering SHG selectively sensitive to non-centrosymmetric biological structures such as collagen fibrils, myosin filaments, and microtubules[137,139,140]. THG, by contrast, is a third-order nonlinear optical process that can occur in both centrosymmetric and non-centrosymmetric materials[141]. In tightly focused microscopy, THG from homogeneous regions is often weak because generated fields tend to cancel through Gouy-phase-related phase mismatch, whereas optical discontinuities break this cancellation and produce strong interface contrast[142]. As a result, THG signals are preferentially generated at optical interfaces and in regions with abrupt changes in refractive index or third-order nonlinear susceptibility[143,144].

Quantitative analysis of SHG and THG images yields parameters directly relevant to pathological assessment. In SHG imaging, commonly extracted metrics include collagen-fibre orientation, length, width, and density, as well as higher-order organisational descriptors such as coherency, entropy, and texture statistics[145-148] (Fig. 3c). Polarisation-dependent measurements and forward-to-backward emission ratios further report on collagen packing and hierarchical organisation[146,149]. Taken together, these parameters describe ECM architecture and remodelling[148,150]. In THG imaging, quantitative features include interface density, lipid-droplet size and distribution, spatial signal heterogeneity, and boundary-associated contrast (Fig. 3d). These features reflect RI discontinuities and intracellular structural organisation[143,144,151].

SHG and THG microscopy are useful in pathology because they quantify tissue architecture and microenvironmental organisation without labelling. In oncological pathology, SHG has been extensively applied to characterise collagen reorganisation at tumour-stroma interfaces, where changes in fibre alignment and orientation have been reported in association with tumour invasion and disease progression[150,152,153]. THG complements this structural information by visualising lipid distribution and cellular interfaces within the tumour microenvironment, providing additional insight into metabolic and structural heterogeneity[143,151,154]. Together, SHG and THG support assessment of tissue architecture in fresh, unprocessed specimens, including applications such as intraoperative evaluation and tumour-margin assessment where conventional staining is impractical[154-156]. Beyond oncology, SHG and THG microscopy address pathological questions related to tissue remodelling and structural integrity. In fibrotic diseases and tissue repair, SHG-derived collagen metrics provide quantitative measures of matrix maturation and organisation[150,157], while THG highlights interface-level changes associated with lipid accumulation or cellular reorganisation[143,144]. These capabilities also extend to tissue engineering and regenerative medicine, where they can be used to monitor scaffold organisation and ECM development[156].

Despite these strengths, SHG and THG microscopy remain constrained by their contrast-specific physical mechanisms. Bulk SHG contrast is primarily limited to non-centrosymmetric structures, such as fibrillar collagen, myosin, and other ordered biomolecular assemblies. THG, by contrast, is not subject to centrosymmetry restrictions, but generates appreciable contrast only at optical interfaces. Such contrast specificity restricts their ability to probe many cellular and molecular features relevant to disease[137,143]. Additionally, imaging depth is typically limited to a few hundred micrometres, and the requirement for ultrafast laser sources and precise optical alignment increases system complexity and cost[136,137,158].

Although AI does not fundamentally overcome the intrinsic physical and hardware limitations of SHG and THG microscopy, it has improved downstream analysis and supported diagnostic performance. In downstream analysis, AI supports quantitative evaluation of collagen architecture and interface-sensitive features related to tissue remodelling. Deep-learning-based segmentation methods can automate identification of collagen-rich regions, regions of aligned fibres, and tissue boundaries from SHG images[145,159,160]. Encoder-decoder architectures have been applied to virtually stain SHG and multiphoton datasets, enabling direct comparison with conventional histological sections[161]. For diagnostic performance, machine-learning models trained on SHG- and THG-derived features have demonstrated classification performance in cancer detection, fibrosis staging, and tissue classification[159,160,162,163]. These approaches typically operate on physically interpretable structural features, preserving the link between optical contrast and pathological meaning[13,164].

Having examined how refractive-index, polarisation, and nonlinear contrasts encode morphological and structural organisation, we next turn to a physically distinct information axis: the mechanical response of cells and tissues.

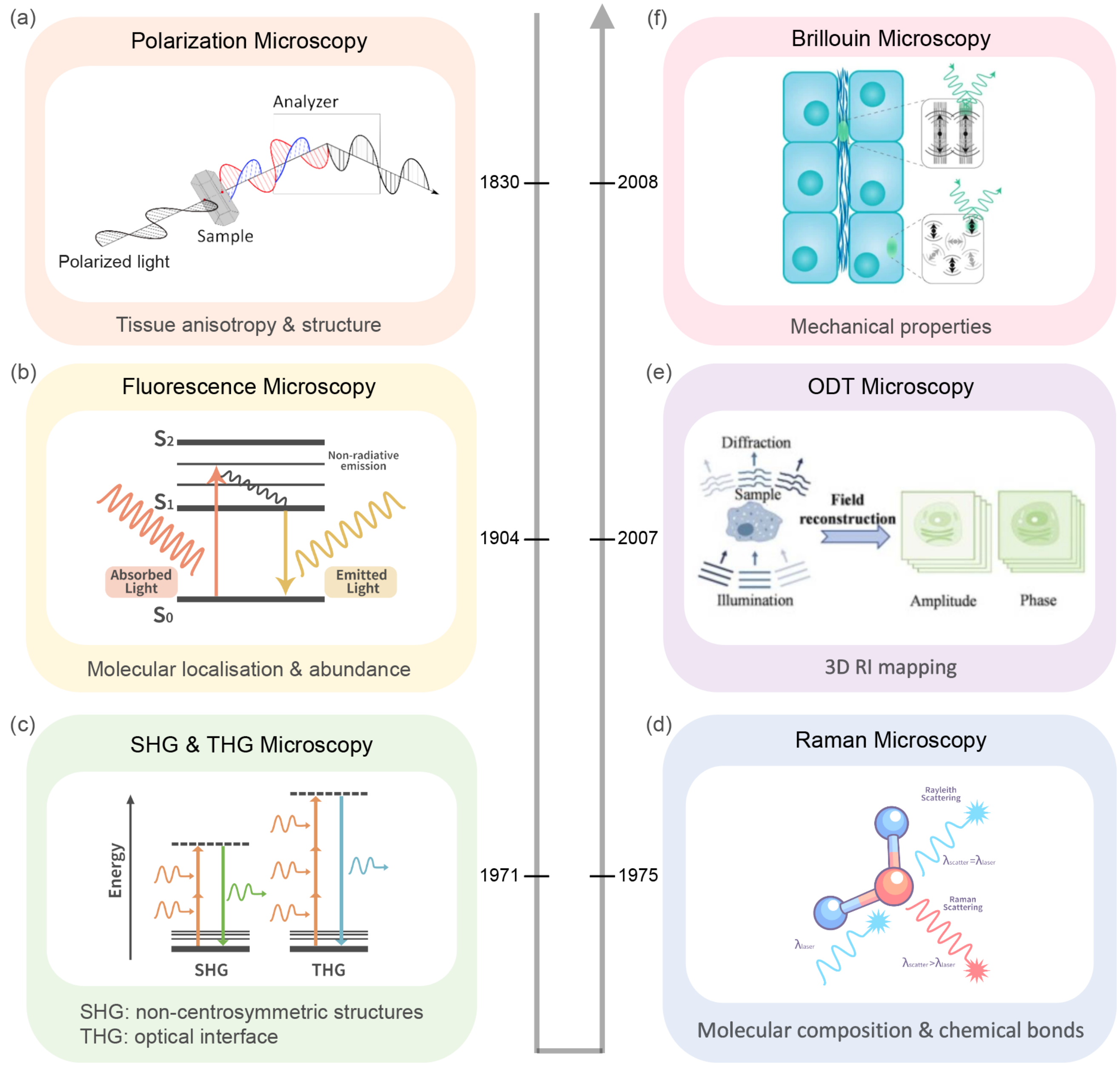

**Fig. 2. Timeline of representative optical microscopy modalities.** Each panel illustrates the imaging principle and the key physical parameters probed. a, Polarisation microscopy[165]; b, Fluorescence microscopy; c, SHG and THG microscopy; d, Raman microscopy; e, ODT[70]; f, Brillouin microscopy[166]. (a) Adapted from an image by Rhian Patterson (ref. [165]), licensed under CC BY-SA 4.0. Changes were made. (e) Adapted with permission from ref. [70]. © Elsevier Ltd. (f) Adapted with permission from ref. [166]. © Springer Nature.

## 2.2 Mechanical properties

Mechanical properties constitute a distinct and complementary dimension of pathological information, reflecting how cells and tissues respond dynamically to deformation[167]. From an optical perspective, mechanical contrast probes material responses that arise from cytoskeletal organisation, ECM composition, and tissue hydration, integrating multiple aspects of disease-associated remodelling into a single physical readout[168]. In this section, we focus on Brillouin microscopy as a representative optical approach for probing mechanical properties, emphasising how high-frequency elastic and viscoelastic responses complement structural and molecular information in pathological imaging.

### *2.2.1 Brillouin microscopy*

Brillouin microscopy probes the mechanical properties of biological samples by measuring inelastic light scattering from thermally excited acoustic phonons[166,169,170] (Fig. 2f). The scattering produces frequency-shifted Brillouin peaks in the scattered spectrum. The frequency shift is related to the real part of the longitudinal modulus and reflects elastic properties at gigahertz probing frequencies, while the Brillouin peak linewidth is associated with viscous dissipation[171,172].

From an implementation perspective, Brillouin microscopy can be realised using either spontaneous or stimulated scattering schemes[173,174]. Confocal Brillouin microscopy enabled by virtually imaged phased-array spectrometers greatly improved acquisition speed and detection efficiency compared with early scanning-interferometric approaches[173,175]. Subsequent developments such as line-scanning Brillouin microscopy further increased throughput by multiplexing spectral acquisition along one spatial dimension, reducing per-pixel dwell time and irradiance[176-178]. Stimulated Brillouin scattering implementations employ counter-propagating pump and probe beams to coherently drive acoustic waves, providing signal levels typically higher than those of spontaneous scattering[174,179]. More recent schemes, including impulsive stimulated Brillouin scattering and Brillouin gain microscopy, have been developed to address limitations in temporal resolution and phototoxicity, expanding the range of biological samples that can be studied[180-184].

At the parameter level, variations in the Brillouin frequency shift have been correlated with changes in cytoskeletal organisation, intracellular crowding, and solid-liquid composition across different cellular states[185-187]. Although linewidth measurements give access to viscous information, their biological interpretation is often complicated by experimental factors and spatial averaging, so most pathological studies have focused on frequency-shift-based elastic contrast[166,188,189]. We note that the GHz longitudinal modulus measured by Brillouin microscopy is a distinct physical quantity from conventional low-frequency Young's or shear moduli, differing by orders of magnitude, so any correlation with macroscopic tissue mechanics is empirical and sample-dependent[166,169].

Brillouin microscopy is particularly suited to diseases in which mechanical alterations accompany disease onset or progression (Fig. 3e). In ophthalmic pathology, Brillouin imaging has been increasingly applied to characterise corneal biomechanics, revealing changes in Brillouin-derived longitudinal modulus contrast associated with keratoconus, typically as focal, spatially localised softening, and age-related alterations of the crystalline lens[190-192]. In cardiovascular research, Brillouin microscopy has been used to map mechanical heterogeneity in myocardial tissue and atherosclerotic plaques, supporting studies of vascular remodelling and cardiac disease[193,194]. In oncology, Brillouin microscopy has revealed spatial heterogeneity in Brillouin frequency shift within tumours and cancer cell populations, providing insight into biomechanical remodelling associated with malignancy and metastatic potential[195-197]. These measurements have also demonstrated potential value for assisting tumour-margin assessment when combined with structural imaging modalities[195,198]. Beyond soft tissues, Brillouin microscopy has been applied to bone and dental tissues, where differences in Brillouin signatures between healthy and diseased states reflect changes in mineralisation and matrix organisation[199-201]. More recently, Brillouin imaging has been used in 3D bioprinting to non-invasively evaluate the mechanical integrity and functional maturation of printed constructs[202-204].

Despite its promise, routine pathological implementation of Brillouin microscopy remains constrained by both mechanical interpretation and technical implementation. The frequency-dependent nature of the measurement, together with the dependence of the Brillouin shift on local refractive index, density and especially water content, complicates direct mechanical interpretation[166,169,205]. To address this, AI-based multivariate analysis has been introduced to extract reproducible mechanical phenotypes from high-dimensional Brillouin data, supporting classification of pathological states beyond simple frequency-shift mapping[206-208]. Practical limitations further arise from the inherently weak Brillouin signal and limited acquisition speed, which can restrict throughput in large or dynamic specimens[173,174,177]. Machine-learning spectral analysis can help ease these data-processing constraints by speeding up frequency-shift extraction and improving robustness under low-signal conditions compared with conventional peak fitting[206,207]. Beyond addressing data-level challenges, deep-learning models have also been applied to Brillouin imaging of multicellular spheroids and tissue constructs, supporting the evaluation of mechanically derived features as candidate biomarkers[208].

### 2.3 Chemical composition and molecular specificity

After structural and mechanical readouts, chemical composition and molecular specificity provide a third pathological information axis, revealing what biological tissues are made of and how their biochemical states change during disease[15,209]. From an optical perspective, chemical contrast arises from interactions between light and molecular vibrational or electronic states, providing direct access to biochemical composition, molecular concentration, and chemical heterogeneity[210]. Unlike purely structural or mechanical contrasts, chemical and molecular contrasts enable identification of specific tissue constituents and metabolic states that may not manifest as morphological changes[211]. In this section, we focus on Raman microscopy and fluorescence microscopy as representative optical strategies for probing chemical composition and molecular specificity.

#### *2.3.1 Raman microscopy*

Raman microscopy is a vibrational spectroscopic imaging technique that combines optical microscopy with Raman scattering to provide label-free mapping of molecular composition in biological samples[212]. The physical basis of Raman microscopy lies in inelastic light–matter interactions in which incident photons exchange energy with molecular vibrational modes (Fig. 2d). The resulting Raman shift encodes a molecular fingerprint that is characteristic of specific chemical bonds and functional groups[213]. As a result, Raman microscopy provides direct access to biochemical composition rather than structural contrast alone[212,214,215]. Raman spectra typically span two diagnostically important regions: the fingerprint region, which contains rich information on proteins, nucleic acids, lipids, and metabolites[216], and the high-wavenumber C-H stretching region, which is sensitive to lipid content, degree of saturation, and molecular packing[217]. These spectral features provide the basis for semi-quantitative biochemical characterisation in pathological samples[212,218,219].

Analysis of Raman spectra yields semi-quantitative parameters relevant to disease-associated biochemical remodelling[216,218,220-224]. Protein-related bands such as the amide I region encode secondary structure and conformational changes[221], while lipid-associated bands reflect abundance, spatial distribution, and metabolic state[218,222]. Alterations in nucleic-acid-associated features indicate changes in cellular proliferation and transcriptional activity, whereas shifts in metabolic signatures reflect disease-related reprogramming[219,225,226]. In nonlinear Raman modalities such as stimulated Raman scattering (SRS) and coherent anti-Stokes Raman scattering (CARS), coherent vibrational contrast enables rapid chemical mapping of tissue components including lipid droplets, nuclei, and cytoplasmic regions with submicron spatial resolution[227]. Note that unlike SRS, CARS is complicated by a non-resonant background that distorts spectra and hampers quantitative interpretation[214,228-230].

The pathological value of Raman microscopy lies in its ability to reveal biochemical alterations that may precede or accompany morphological change (Fig. 3f)[218,220-223,231,232]. In tissue pathology, Raman imaging has been used to distinguish healthy and diseased tissue states by identifying disease-associated chemical signatures, supporting tumour diagnosis and grading across multiple organ systems[218,220-222]. Chemical sensitivity to lipid-rich regions has demonstrated utility in delineating tumour margins, where metabolic heterogeneity and lipid remodelling are prominent[218,233]. In neurological pathology, Raman spectroscopy has been applied to detect biochemical signatures associated with neurodegenerative diseases, in which protein aggregation and conformational change play central roles[231,232]. In infectious disease and microbiology, Raman microscopy enables rapid label-free identification of bacteria and fungi based on their intrinsic chemical fingerprints[234,235]. This capability supports microbial diagnostics, biofilm characterisation, host-pathogen interaction studies, and assessment of antibiotic response, where chemical composition provides a direct readout of cellular state[234-237].

Despite these strengths, Raman microscopy faces several challenges for routine pathological deployment. The intrinsically weak signal of spontaneous Raman scattering limits imaging speed and imaging depth[214,238]. Although nonlinear Raman techniques such as SRS and CARS can increase signal levels and imaging throughput, they require greater system complexity and higher cost, and CARS additionally complicates quantitative spectral interpretation[214]. Analytical challenges compound these hardware constraints. First, autofluorescence frequently obscures diagnostically relevant vibrational features, complicating spectral interpretation[214,238]. Machine-learning and deep-learning denoising methods can suppress fluorescence contamination and background interference, helping recover weak Raman signatures embedded in noisy spectra[239]. Second, spectral overlap among biochemical constituents hampers reliable feature identification and disease discrimination[219,225,226]. Classical multivariate methods such as principal component analysis and linear discriminant analysis have long been employed to reduce dimensionality and extract disease-associated spectral features[240,241]. More recently, deep-learning models have been used to help disentangle overlapping vibrational signatures through high-dimensional feature learning, supporting cancer detection and tissue classification performance[242-247]. Third, biological heterogeneity introduces inter- and intra-sample spectral variability, further complicating reproducible interpretation[219,225,226]. Learning-based approaches trained on large, diverse datasets can improve robustness to such variability and support more scalable and consistent pathological assessment[242-246].

### *2.3.2 Fluorescence microscopy*

In contrast to Raman microscopy, which achieves label-free chemical specificity through vibrational spectra, fluorescence microscopy typically uses targeted molecular labelling, or endogenous autofluorescence in label-free variants, to visualise selected cellular and molecular components by detecting light emitted from fluorescent molecules within biological samples[248-250]. Its central importance in pathology arises from its ability to provide direct molecular specificity, enabling visualisation of disease-associated biomarkers that cannot be resolved through morphology alone. Advances in fluorophore chemistry, optical instrumentation, and imaging strategies have established fluorescence microscopy as a foundational modality in modern pathological research and diagnostic workflows[251].

The physical basis of fluorescence microscopy lies in electronic excitation and radiative relaxation of fluorescent molecules (Fig. 2b)[248,252,253]. Fluorophores absorb photons at an appropriate excitation wavelength, transition to an excited electronic state, and subsequently emit lower-energy photons as they return to the ground state. This spectral separation between excitation and emission allows efficient background suppression using optical filters. As a result, fluorescence microscopy enables selective visualisation of labelled biomolecules within complex biological environments. Unlike many label-free modalities that infer biological properties indirectly, fluorescence microscopy directly reports molecular presence, localisation, and interaction, making it well suited for pathology driven by molecular dysregulation[254].

Fluorescence microscopy is not limited to intensity-based information; it also provides access to a broad range of quantitative and functional parameters. Fluorescence intensity reflects relative biomarker abundance, while spatial distribution reveals subcellular localisation and tissue architecture[255-257]. Time-resolved fluorescence measurements enable investigation of dynamic processes such as vesicle trafficking, receptor internalisation, and cell-cell interactions[258,259]. Fluorescence lifetime imaging microscopy, being largely independent of fluorophore concentration, adds sensitivity to the molecular microenvironment, binding state, and metabolic activity, while fluorescence anisotropy imaging reports on molecular mobility and protein interactions[260]. Together, these contrasts, from lifetime to anisotropy, enable multiparametric characterisation of pathological states beyond conventional histological staining.

Advances in optical design have further expanded the accessible parameter space of fluorescence microscopy. Light-sheet fluorescence microscopy enables rapid volumetric imaging of large specimens with reduced photodamage, providing 3D spatial context across extended tissue volumes[16-18]. Multiphoton excitation microscopy allows deeper tissue imaging with reduced out-of-focus phototoxicity and photobleaching, supporting 3D imaging of intact or living samples[22,136,261]. Super-resolution techniques such as structured illumination microscopy, stimulated emission depletion microscopy, and single-molecule localisation microscopy overcome the diffraction limit of light and enable visualisation of nanoscale pathological features including receptor clustering, cytoskeletal remodelling, and protein aggregation[9,19-21,262-265].

Fluorescence microscopy is central to pathological applications that require molecular specificity and targeted interrogation of disease mechanisms (Fig. 3g). In oncological pathology, fluorescence imaging has been widely used to visualise tumour-associated biomarkers, assess intratumoural heterogeneity, and delineate tumour margins through multiplexed labelling strategies[266-268]. Immunofluorescence enables precise localisation of oncogenic signalling proteins, immune markers, and therapeutic targets, supporting tumour classification, prognostic evaluation, and treatment selection[269,270]. Advanced fluorescence approaches further facilitate characterisation of the tumour microenvironment, including immune cell infiltration, stromal organisation, and angiogenic processes[271,272]. In neurological pathology, fluorescence microscopy is widely used to study neurodegenerative diseases by visualising pathological protein aggregates, synaptic alterations, and glial responses[273]. In infectious disease pathology, fluorescence-based methods enable rapid detection and identification of pathogens and support visualisation of host-pathogen interactions and biofilm formation[274].

Despite these strengths, fluorescence microscopy faces several limitations that constrain routine pathological deployment. First, photobleaching and phototoxicity restrict long-term or live imaging, and imaging depth remains limited by light scattering in thick tissues. These constraints arise primarily from photophysical and optical properties and cannot be directly resolved through computational approaches. Second, variability introduced by labelling strategies and sample preparation complicates quantitative comparison across specimens and laboratories. AI-based segmentation and feature-extraction methods can enable automated and standardised identification of cellular and subcellular structures across large fluorescence datasets[275-277]. Learning-based models can further improve robustness under incomplete, variable, or heterogeneous labelling, giving more consistent detection of pathological regions and phenotypes[278-280]. Third, multiplex fluorescence imaging generates high-dimensional datasets in which signals from multiple channels must be interpreted jointly. AI can facilitate multiplex analysis by integrating information across fluorescence channels and, when applicable, imaging modalities, enabling quantitative characterisation of complex molecular landscapes[279,280]. Finally, advanced fluorescence modalities such as super-resolution and multiphoton microscopy increase system complexity and data volume. Machine-learning approaches can accelerate image reconstruction, denoising, and downstream analysis, reducing manual annotation burden and improving throughput[275,277,281-290].

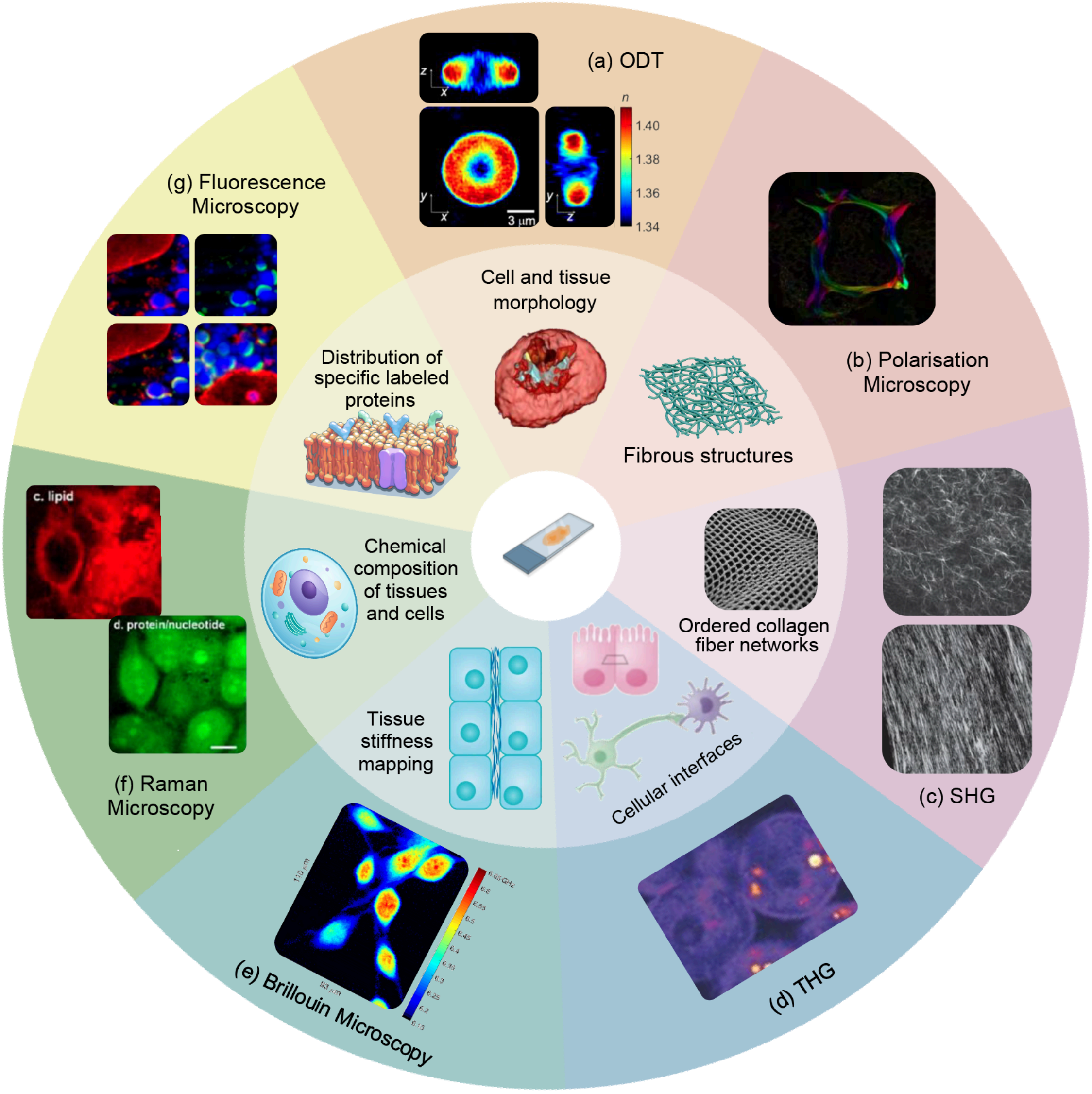

**Fig. 3. Modality-specific sensitivity and disease relevance of seven representative optical microscopy techniques for pathological imaging.** The inner layer indicates the biological structures or physical properties to which each modality is most sensitive; the outer layer illustrates the characteristic imaging appearance typically produced by each technique. a, ODT, sensitive to 3D RI distributions, enabling quantitative analysis of cellular morphology and mass-density-related phenotypes; relevant to haematological disorders, infectious diseases such as malaria, and single-cell phenotyping in cancer[32]. b, Polarisation microscopy, sensitive to birefringent and anisotropic fibrous structures such as aligned collagen networks; widely applied in fibrosis staging, evaluation of desmoplastic tumour stroma, and quantitative assessment of breast, colorectal, and cervical cancers. c, SHG microscopy, structure-specific nonlinear imaging of non-centrosymmetric fibrillar collagen[137,139]; used to assess tumour-stroma interfaces, fibrotic remodelling, and ECM reorganisation in breast, skin, and cardiovascular diseases[146]. d, THG microscopy, sensitive to refractive-index discontinuities and lipid-rich cellular interfaces[143,144]; applied in skin cancer, neurological disorders, lipid-associated metabolic diseases, and infection-related structural alterations[143]. e, Brillouin microscopy, sensitive to high-frequency mechanical contrast and biomechanical heterogeneity[166]; applied to ophthalmic diseases, cardiovascular remodelling, bone pathology, and tumour characterisation[192,193,195,200]. f, Raman microscopy, probing intrinsic molecular vibrational signatures for label-free biochemical characterisation of proteins, lipids, and nucleic acids; demonstrated utility in tumour diagnosis and grading (lung, breast, colon, prostate), neurodegenerative diseases, microbial identification, and metabolic disorder assessment[291]. g, Fluorescence microscopy, providing molecular specificity through targeted labelling of biomarkers and enabling multiplexed molecular interrogation; central to oncological biomarker detection, immune microenvironment profiling, neurodegenerative pathology, and infectious disease diagnostics[248]. (a) Adapted

with permission from ref. [32]. CC BY 4.0. (c, inner layer) Adapted with permission from ref. [292]. CC BY 4.0. (c, outer layer) Adapted with permission from ref. [146]. © Springer Nature. (d, inner layer) Adapted with permission from ref. [293]. © Springer Nature. (d, outer layer) Adapted with permission from ref. [143]. © Nature Publishing Group. (e, inner layer) Adapted with permission from ref. [70]. © Elsevier. (e, outer layer) Adapted with permission from ref. [183]. © Optical Society of America. (f, outer layer) Adapted with permission from ref. [291]. © American Chemical Society. (g, outer layer) Adapted with permission from ref. [294]. © The Company of Biologists.

## 3. Modality selection for pathology-specific imaging tasks

Optical microscopy modalities differ substantially in their operational characteristics, so choosing among them is a matter of fitness for purpose rather than of finding a single technique that is universally optimal across all pathological contexts. In practice, the best modality is determined by the diagnostic task and its required balance of imaging speed, spatial scale, molecular specificity, interpretability and workflow compatibility (see Table 1 for a summary of operational parameters and workflow considerations across all seven modalities). In the following subsections, we match representative modalities to several common pathological scenarios, drawing out the practical trade-offs that shape modality selection in real diagnostic workflows.

### 3.1 Intraoperative assessment and rapid decision-making

For time-sensitive pathological tasks, including intraoperative frozen-section evaluation, tumour-margin assessment, and endoscopic screening, the main constraints are imaging speed, ease of interpretation and minimal sample preparation[5,295-297]. These requirements favour modalities that can rapidly survey tissue architecture and generate visually intuitive contrast for pathological interpretation.

Polarisation microscopy, particularly in modern snapshot implementations, is advantageous because it can assess stromal organisation and collagen-related architectural disruption over large tissue areas without staining[298]. Fluorescence microscopy also suits rapid workflows when autofluorescence or fast vital staining is used, giving molecularly informative signals whose visually interpretable contrast aligns with established pathological descriptors[299,300].

By comparison, modalities that rely on slower point-scanning acquisition or computationally intensive reconstruction sit less naturally within time-sensitive workflows. SHG and THG microscopy remain limited by instrumentation complexity and restricted clinical accessibility, which hinder their widespread intraoperative implementation[296]. Brillouin and conventional Raman microscopy face even stronger constraints in this setting because slow spectral acquisition and indirect parameter interpretation can substantially delay clinical decision-making[170,222].

These trade-offs illustrate a recurring tension in pathological imaging: modalities optimised for rapid large-area assessment often sacrifice quantitative specificity, whereas quantitative modalities are frequently limited by throughput and interpretability constraints.

### 3.2 Large-area tissue assessment and stromal remodelling

When pathological interpretation depends on tissue-wide organisation, such as in fibrosis staging or stromal remodelling assessment, a large field of view and tissue-scale spatial context matter more than localised quantitative sensitivity. Such tasks therefore favour modalities that preserve mesoscale organisation across extended tissue regions.

Polarisation microscopy is particularly compatible with this requirement because polarisation-derived contrast can reveal collagen alignment and stromal heterogeneity over millimetre-scale fields of view[93,94,98,107,108]. Additionally, SHG microscopy provides a complementary view by enabling quantitative characterisation of fibrillar collagen and ECM architecture at the local tissue scale, over limited fields of view, whereas polarisation microscopy provides a broader, large-area overview of stromal organisation[145-148,150].

By contrast, modalities optimised for localised quantitative measurements, such as Raman microscopy[214,238], Brillouin microscopy[173,174,177], and ODT[43,46-48], are generally less practical for whole-specimen architectural evaluation. Their smaller imaging regions and slower acquisition limit scalability across large pathological sections, again reflecting the trade-off between large-area assessment and localised quantitative specificity noted in the previous subsection.

### 3.3 Single-cell phenotyping and mechanistic insight

Applications centred on single-cell analysis, such as haematological malignancies, circulating tumour cell characterisation, and infectious disease assessment, place greater emphasis on spatial resolution, quantitative sensitivity, and multidimensional parameter extraction than on imaging throughput or large-area coverage. These requirements favour modalities capable of resolving subtle variations in cellular morphology, intracellular organisation, biomechanical state, or biochemical composition with high spatial detail.

ODT is well suited because label-free RI-derived measurements enable quantitative characterisation of cellular morphology and intracellular heterogeneity with subcellular resolution, although dry-mass-related quantities rest on an assumed specific refractive increment. Other modalities can extend this analysis beyond structure: Raman microscopy provides complementary biochemical specificity through intrinsic vibrational signatures[214,216,217], while Brillouin microscopy contributes mechanical information associated with intracellular biomechanical heterogeneity[185,186,195-197].

These practical strengths come with real limitations: the outputs often require computational interpretation and align less readily with conventional morphology-based pathological descriptors[46,47,55,166,169,214,219]. In addition, the acquisition strategies required for high-dimensional quantitative measurements frequently reduce throughput and complicate integration into routine diagnostic workflows[43,46-48,173,174,177,214,238]. Therefore, these modalities currently occupy a space closer to mechanistic insight and quantitative phenotyping than to high-throughput clinical pathology, where scalability and interpretability remain dominant operational requirements.

### 3.4 3D tissue organisation and volumetric pathology

When pathological processes extend across multiple tissue layers, such as tumour invasive fronts, fibrotic septa in intact tissues, and 3D organisation within organoids, imaging depth and optical sectioning become the key requirements. These requirements favour modalities capable of preserving 3D structural context within intact or minimally processed specimens.

Multiphoton fluorescence microscopy, together with SHG and THG microscopy, is particularly advantageous because nonlinear optical sectioning enables volumetric imaging while maintaining tissue architecture at moderate subsurface depths[22,154-156]. This matters when evaluating invasive tumour fronts, ECM remodelling, and volumetric tissue organisation that thin two-dimensional sections cannot adequately represent[150-152,154-157]. In practice, however, their use can be limited by optical complexity, acquisition time and potential phototoxicity. ODT also provides label-free volumetric imaging of pathological tissues, although multiple scattering remains a challenge for clinical applications[46,48]. Light-sheet fluorescence microscopy further enables large-volume 3D imaging of cleared tissues, although extensive sample preparation limits its clinical applicability[301,302].

Conversely, modalities optimised for superficial imaging or thin-section analysis, such as conventional wide-field fluorescence imaging and polarisation microscopy, are generally less effective for volumetric pathological assessment because these wide-field methods lack intrinsic optical sectioning; scattering and depth-dependent signal degradation further restrict reliable imaging of large 3D structures[46,47,303].

This scenario therefore illustrates another central trade-off in optical pathology: achieving volumetric structural information often comes at the cost of greater optical complexity and lower imaging throughput compared with conventional two-dimensional workflows.

### 3.5 Molecular phenotyping and metabolic assessment

When pathological changes are driven by or reflected in molecular alterations, metabolic states or biomarker expression, such as in tumour metabolic profiling, microbial identification, immune microenvironment characterisation and protein-aggregation imaging in neurodegenerative disease, biochemical and molecular specificity become the dominant considerations. These requirements favour modalities capable of directly probing molecular composition, metabolic heterogeneity, or biomarker expression.

Raman microscopy is particularly advantageous for label-free biochemical characterisation because intrinsic vibrational signatures give access to lipid, protein, and, with sufficient signal, metabolite and nucleic-acid composition without exogenous labelling[212,214,216,218]. Alternatively, fluorescence microscopy can access the same pathological dimension through targeted labelling strategies, while offering higher throughput and established multiplex imaging workflows[248-251,266-268].

These two approaches differ in their use of intrinsic and targeted contrast mechanisms. Raman microscopy preserves label-free biochemical specificity but remains constrained by weak signal intensity and relatively slow spectral acquisition[214,238]. Exogenously labelled fluorescence microscopy instead sacrifices label-free operation in exchange for higher sensitivity, faster imaging, and stronger compatibility with established pathological workflows[248-250]. In both cases, increasingly high-dimensional datasets create a growing dependence on computational analysis and AI-assisted interpretation, especially when multimarker or multimodal information must be integrated in complex tissue.

**Table 1. Summary of imaging characteristics of seven optical microscopy modalities for pathology.** The table summarises the performance characteristics and workflow considerations of the microscopy modalities discussed above, including spatial resolution, imaging speed, imaging depth, sample compatibility, interpretability, and workflow complexity. Quantitative values indicate representative order-of-magnitude ranges reported in commonly adopted implementations rather than optimised, prototype-level, or emerging demonstrations that have not yet been systematically validated. Speed values are reported in modality-native units and are not directly comparable across point-scanning, frame-based, and volumetric implementations. Spatial resolution values reflect lateral resolution; axial resolution is not listed because it is generally lower and varies with numerical aperture, modality, and specific implementation. The roles of AI described here refer specifically to modality-dependent applications within each microscopy technique. Broader functions of AI are discussed in the Outlook section.

| # | ODT | Polarisation microscopy | SHG & THG | Brillouin microscopy | Raman microscopy | Fluorescence microscopy |
|---|---|---|---|---|---|---|
| **Imaging speed / throughput** | Seconds per 3D volume[304] (moderate) | Conventional: ~10 s–min per image (low)[113]; Snapshot: ms–s per image (moderate)[83,305] | ~1–10 s per image[155] (moderate) | ~10–100 ms per pixel[177] (low) | Raman: ~0.1–1 s per pixel (low); SRS: video-rate[214] (moderate) | Video-rate to high-throughput[248 16-18] (high) |
| **Imaging depth** | ~10–100 µm[32] (moderate) | ~100–500 µm[306] (moderate) | ~200–500 µm[136,155](moderate) | ~50–200 µm[169] (moderate) | ~10–100 µm[214,219] (superficial–moderate) | Widefield/confocal : ~10–100 µm[136,307] (superficial–moderate); MPM: ~100–500 µm[303] (moderate) |
| **Field of view** | ~50–200 µm[32] (small–moderate) | mm–cm scale[308] (large-area) | ~200–500 µm[137] (moderate) | ~50–300 µm[166] (small-moderate) | ~10–100 µm[238](small–moderate) | mm–whole-slide[5] (large-area) |
| **Spatial resolution (lateral)** | ~200–500 nm[32,308] (subcellular) | ~0.5–2 µm effective[309] (low-NA /tissue microstructure) | ~300–500 nm[143] (subcellular–tissue) | ~300–800 nm[177] (subcellular) | ~300–800 nm[230] (subcellular) | ~20–300 nm (SR)[262-264]; ~200 nm (confocal)[248] |
| **Sample compatibility** | Live cells; thin tissue sections[32] | Tissue sections; bulk tissue; endoscopic[114,310] | Fresh tissue; thick tissue; ex vivo[311] | Live cells; soft tissues; ex vivo[177] | Fresh tissue; sections; microorganisms[312] | Fixed sections; live cells[248,249] |
| **Interpretability** | Model-dependent[43,304] | Requires parameter decomposition[97] | Structure-specific[313] | Indirect mechanical meaning[166,169] | Spectral-model dependent[212,219] | Directly interpretable |
| **Workflow complexity** | Tomographic acquisition and reconstruction[304] (Moderate) | Parameter decomposition and calibration[97] (Moderate) | Ultrafast laser system and alignment[156,314] (High) | Spectral acquisition and interpretation[185] (High) | Spectral acquisition and analysis[315] (High) | Labelling and staining workflow[248,250] (Moderate) |
| **Key practical limitations** | Multiple scattering; indirect RI interpretation[32,43] | Geometry- and scale-dependent parameters[71] | Limited molecular specificity; complex optics[316] | Complex interpretation; slow acquisition[166,317] | Weak signal; autofluorescence; spectral overlap among biochemical constituents; limited imaging depth[214,219,318] | Photobleaching; phototoxicity; labelling variability; scattering-limited imaging depth[248,250,303] |
| **Imaging properties** | Quantitative structural backbone | ECM / anisotropy context | ECM / interface contrast | Mechanical complement | Chemical specificity | Molecular specificity |
| **Role of AI** | Reconstruction; parameter-space analysis[304,319] | Parameter extraction; feature-based classification[129,320] | Structural segmentation; pattern quantification[159,160,162,163] | Spectral fitting; mechanical phenotype analysis[206,207] | Spectral interpretation; biochemical feature extraction[240-242,244-246] | Segmentation; multiplex channel analysis; molecular quantification[278-280] |

## 4. Discussion

Advanced optical microscopy adds pathological information beyond conventional morphology-based histology because its quantitative readouts carry complementary structural, mechanical and biochemical information about tissue. This review provides a pathology-centred overview of advanced optical microscopy, linking optical principles and quantitative readouts to pathological applications. A central perspective is task-guided modality selection: different pathological questions require different balances of contrast, scale, speed, interpretability and workflow compatibility. Together, these discussions lead to an outlook along three connected directions: improving individual modalities, developing multimodal strategies, and deepening the integration of AI.

At the level of individual modalities, future development can be considered in terms of four key aspects: accessibility, throughput, interpretability and imaging fidelity (Fig. 4a). First, accessibility can be improved through affordable, portable and user-friendly system designs, such as compact interferometric configurations, simplified light-sheet geometries and modular optical platforms[32,301,321-324]. Second, throughput can be improved through faster reconstruction algorithms, curated databases, and single-shot or minimally multiplexed strategies, reducing the latency from imaging to diagnosis[44,150,325]. Third, interpretability can be strengthened by establishing clearer links between optical parameters, biophysical properties and pathological descriptors[86,150,326]. This will require closer collaboration from laboratory development to clinical pathology. Fourth, imaging fidelity can be improved by using adaptive optics to compensate for tissue-induced aberrations. By correcting these aberrations, it can improve signal recovery, preserve contrast fidelity and extend effective imaging depth in optically heterogeneous pathological specimens[26,327-335]. Together, these four aspects show real potential for advanced optical microscopy in pathology, while pointing to the challenges that remain for routine clinical translation.

At the level of multimodal microscopy, future development can be considered through two main strategies: cross-validating pathological features when modalities provide overlapping information, and combining complementary readouts when modalities capture distinct information (Fig. 4b). First, when information domains partially overlap, multimodal imaging cross-validates shared features[336]. For example, structural organisation measured by ODT or SHG can be interpreted together with fluorescence data, reducing interpretational ambiguity and improving diagnostic robustness[337,338]. Second, for non-overlapping domains, multimodal strategies provide genuinely complementary information, such as combining SHG-based collagen architecture with fluorescence phenotyping[150,152] or integrating Raman-derived biochemical signatures with structural imaging[66,339,340]. Multimodal strategies are therefore attractive, but their clinical translation will require balancing diagnostic benefit against added complexity and cost.

At the level of AI integration, future progress can be considered across two levels: image-level enhancement and pathology-level analysis. Image-level enhancement covers image resolution enhancement, virtual staining, and denoising (left panel of Fig. 4c). First, for resolution enhancement, UNet-based and GAN-based frameworks have been widely used to recover high-quality images from constrained acquisition conditions[282-288]. For example, content-aware restoration and structured-illumination reconstruction networks enable high-fidelity recovery from low-SNR fluorescence images[284,286]. GAN-based cross-modality super-resolution models learn structural mappings from wide-field to super-resolution data and generate inferred high-resolution estimates[287-289]. Second, for contrast generation, GAN-based frameworks have demonstrated conversion of label-free modalities into synthetic H&E or special stains[341-348], while diffusion-based generative models may further improve the stability and fidelity of image synthesis[281,349,350]. However, AI-generated histological images may contain inaccurate or artificial features, requiring careful validation before clinical use[351]. Third, for denoising, residual attention networks and self-supervised learning strategies have been employed to suppress shot noise and acquisition artefacts under low-signal conditions[282,283,285,286], improving quantitative stability across heterogeneous samples.

At the pathology level, AI can support segmentation, diagnostic classification, and predictive modelling (right panel of Fig. 4c)[352]. First, for segmentation, encoder-decoder architectures such as UNet and related convolutional models are used to delineate nuclei, cells, ECM structures, or tumour boundaries from complex tissue backgrounds[161,275,276]. Second, for diagnostic classification, convolutional neural networks and multimodal deep-learning frameworks have been applied across fluorescence, Raman, Brillouin, ODT, and multiphoton imaging to distinguish malignant from benign tissue or to stage disease progression[13,159,160,162,163,207,208,240-246,278-280,353-355] . These models extract multiscale morphological, mechanical, or biochemical features to support pathological decision-making. Third, for predictive modelling, AI models integrate imaging-derived features with pathological or molecular variables to predict disease progression or therapeutic outcomes, as demonstrated in multiphoton fibrosis staging and cancer prognosis studies[162,353]. As these tasks mature, AI is expected to further expand the role of optical microscopy in pathology.

These three directions are not separate: better individual modalities give multimodal strategies more reliable inputs to combine, and AI could strengthen both. By making optical microscopy techniques more accessible, faster and easier to interpret, these advances will support task-appropriate modality selection and help move them towards routine quantitative use alongside histopathology. We hope this review helps guide that transition and encourages further work at the interface of optical microscopy and pathology.

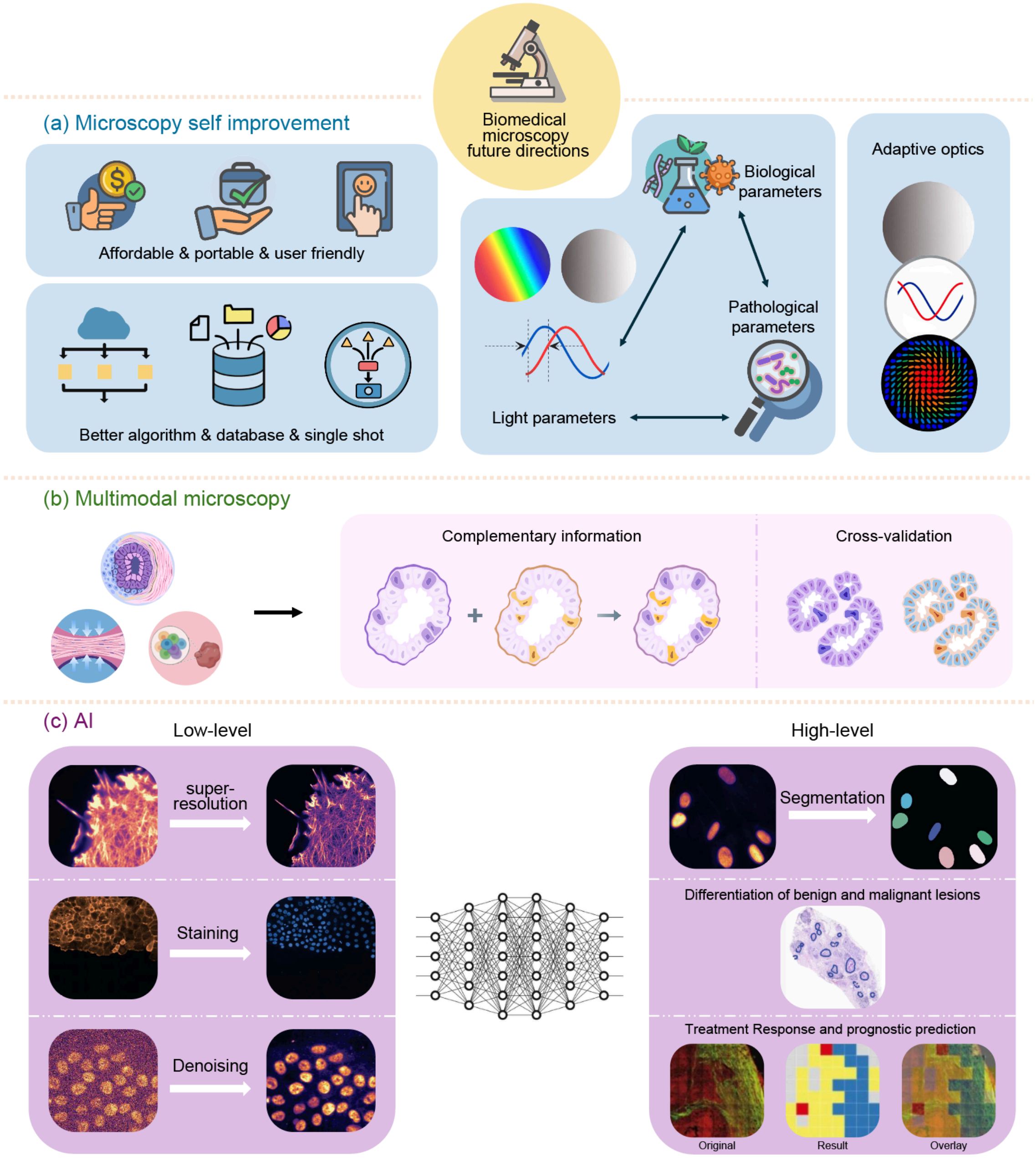

**Fig. 4. Future directions of optical microscopy for pathological applications.** a, Future directions for individual microscopy modalities, highlighting accessibility, throughput, interpretability and imaging fidelity. b, Future directions for multimodal microscopy, highlighting cross-validation from overlapping readouts and complementarity from distinct readouts. c, Future directions for AI integration, highlighting two levels of development: image-level enhancement, including resolution enhancement, contrast generation and denoising (left panel[285,341]), and pathology-level analysis, including segmentation, diagnostic classification and predictive modelling (right panel[4,353]). (a) Adapted with permission from ref. [14]. CC BY 4.0. (c, left panel and upper part of the right panel) Adapted with permission from ref. [14]. CC BY 4.0. (c, middle part of the right panel) Adapted with permission from ref. [4]. © Springer Nature. (c, lower part of the right panel) Adapted with permission from ref. [353]. © Springer Nature.

## Acknowledgements

This work is supported by the Royal Society University Research Fellowship (URF\R1\241734) (C.H.) and the National Natural Science Foundation of China (NSFC) Grant No. 62335007 (H.H.).

## Author Contributions

All authors reviewed the results and approved the final version of the manuscript.

**Competing Interests**

The authors declare no competing interests.

**Additional Information**

Correspondence and requests for materials should be addressed to H.H. or C.H.